\documentclass[journal,onecolumn]{rmaa}

\usepackage{paralist}

\usepackage{psfrag,color}

\usepackage[latin1]{inputenc}
\usepackage{epsfig}
\usepackage{amsmath}
\usepackage{graphicx}
\usepackage{subfig}
\usepackage{multirow}
\usepackage{array}
\usepackage{rotating}

\title{Multicolor study of V1009 Per, a close binary system at the beginning of the overcontact phase, and of CRTS J031642.2+332639, a new binary system in the same field}

\author{
  Ra\'ul Michel\altaffilmark{1},
  Francesco Acerbi\altaffilmark{2},
  Carlo Barani\altaffilmark{2} and
  Massimiliano Martignoni\altaffilmark{2}.
}

\altaffiltext{1}{Instituto de Astronom\'ia, Universidad Nacional Aut\'onoma de M\'exico}
\altaffiltext{2}{Stazione Astronomica Betelgeuse, Magnago Italy}

\shortauthor{Michel et al.}
\shorttitle{Multicolor study of V1009 Per and CRTS J031642.2+332639}

\fulladdresses{
\item
  R. Michel: Universidad Nacional Aut\'onoma de M\'exico.
  Observatorio Astron\'omico Nacional. Apartado Postal 877, C.P. 22800,
  Ensenada, B.C., M\'exico. (rmm@astro.unam.mx)
\item
  F. Acerbi
  Via Zoncada 51, Codogno,
  LO, 26845, Italy. (acerbifr@tin.it)
\item
  C. Barani
  Via Molinetto 35, Triulza di Codogno,
  LO, 26845, Italy. (cvbarani@alice.it)
\item
  M. Martignoni
  Via Don G. Minzoni 26/D, Magnago,
  MI, 20020, Italy. (massimiliano.martignoni@alice.it)
}

\listofauthors{R. Michel, F. Acerbi, C. Barani \and M. Martignoni}
\indexauthor{Michel, R.}
\indexauthor{Acerbi, F.}
\indexauthor{Barani, C.}
\indexauthor{Martignoni, M.}

\abstract{
  The first multicolor observations, and light curve solutions, of the eclipsing binary systems V1009~Per and CRTS~J031642.2+332639 are presented. Using the 2005 version of the Wilson-Devinney code, both systems are found to be W UMa contact binaries. V1009~Per has a mass ratio of $q = 0.362 \pm 0.002$ and a shallow fill out parameter of $f = 11.8 \pm 0.6\%$ while CRTS~J031642.2+332639 has a mass ratio of $q=2.507 \pm0.006$ and a fill out of $f= 13.6 \pm0.4\%$. High orbital inclinations, $i = 85^\circ.9$ for V1009 Per and $i = 83^\circ.2$ for CRTS~J031642.2+332639, implies that both systems are total eclipsing binaries and that the photometric parameters here obtained are reliable. Based on 16 times of minimum, the orbital period variations of V1009~Per are discussed.
  The absolute dimensions of the systems are estimated and, from the logM-logL diagram, it is found that both components of  the systems follow the general pattern of the W subtype W Ursae Majoris systems.}

\resumen{Se presentan las primeras observaciones multicolor, y ajustes a las curvas de luz, de los sistemas binarios eclipsantes V1009~Per y CRTS~J031642.2+332639. Usando la versi\'on 2005 del c\'odigo Wilson-Devinney, se encuentra que ambos sistemas son binarias en contacto del tipo W~UMa. V1009~Per tiene un cociente de masas de $q = 0.362 \pm 0.002$ y un factor de llenado de $f = 11.8 \pm 0.6\%$ mientras que CRTS~J031642.2+332639 tiene los valores de $q=2.507 \pm0.006$ y $f= 13.6 \pm0.4\%$. Las altas inclinaciones orbitales, $i = 85^\circ.9$ para V1009~Per e $i = 83^\circ.2$ para  CRTS~J031642.2+332639, implican que ambas son binarias totalmente aclipsantes y que los parametros fotom\'etricos obtenidos son confiables. En base a 16 tiempos de m\'inimo, se discuten las variaciones del periodo orbital de V1009~Per. Tambi\'en se calculan las dimensiones absolutas de los sistemas y, con el diagrama logM-logL, se encuentra que ambos componentes de los sistemas siguen el patr\'on general del subtipo W de las W Ursae Majoris.}

\addkeyword{techniques: photometric}
\addkeyword{binaries: eclipsing}
\addkeyword{stars: individual: V1009 Per}
\addkeyword{stars: individual: CRTS J031642.2+332639}
\begin{document}
\maketitle

\section{Introduction}
\label{sec:intro}
{The creation of a stellar evolutionary scheme requires a knowledge of the fundamental parameters of stars in different stages of their evolution. Eclipsing binary systems, especially the W Ursae Majoris type, are among the most important sources of such information \citep{Kjurkchieva2017}.}

{W UMa systems have orbital periods, typically, between 0.2 days and 0.8 days and consists of two dwarf stars with spectral types ranging from A to K sharing a common convective envelope resulting in a near equalization of the surface temperature with differences not more than a few percent \citep{Christopoulou2011}.}

{Light curves of W UMa stars show continuous changes in brightness with nearly equal depth minima and maxima that are not always symmetric. This difference in maximum light levels, sometimes referred to as the O'Connell effect \citep{OConnell1951}, is caused by the inhomogeneity in the surface brightness distribution, on one or both stars, commonly associated with dark or hot spots. The difference between the maxima can change from orbit to orbit because of the motion and evolution of these active regions. This phenomenon may indicate the presence of an activity cycle similar to that of the Sun \citep{Mitnyan2018}.}

{In our modern understanding, these systems are most likely formed from the moderate close binaries \citep{Chen2016} through either nuclear evolution of the most massive component in the detached phase  or angular momentum evolution of the two component stars within a convective envelope  \citep{Hilditch1988}; \citep{Tutukov2004}; \citep{Yildiz2013}.}

V1009 Per (GSC 2344-00092, NSVS 6662264, $\alpha_{2000}=03^h16^m49^s.62, \delta_{2000}=+33^\circ30^{'}14^{''}.1$) was first reported by  \citet{Kuruslov2011} as a short-period contact eclipsing binary candidate with an orbital period of about 0.23414 days. The published light curve presented the typical EW-type behaviour however, the data in its light curve are scattered to some extent, and it is not clear that the light curve shows the O'Connell effect or argue about the features of the light curve at the time of the observations. We found the O'Connell effect in our observed light curves and adopt a bright spot model to interpret it.

CRTS J031642.2+332639 (hereinafter J031642, $\alpha_{2000}=03^h16^m42^s.23, \delta_{2000}=+33^\circ26^{'}39^{''}.0$) is listed as variable star, with a period of 0.3009380 days, in the Catalina Surveys Periodic Variable Star Catalog \citep{Drake2014}. This object also shows the typical behaviour of W UMa type systems.

Because no photometric or spectroscopic studies are found in the literature, the aim of the present study is to analyze our $B$, $V$, $R_c$ and $I_c$ light curves to obtain the first orbital and fundamental parameters of these eclipsing binaries.


\section{Observations}

Photometric observations were carried out at the San Pedro Martir Observatory, on December 16, 2017 and January 18, 2018, with the 0.84-m telescope, a filter-wheel and the \textit{Spectral Instruments 1} CCD detector (a  deep depletion e2v CCD42-40 chip with gain of 1.39 e$^-$/ADU and readout noise of 3.54 e$^-$). The field of view was $7.6^{\prime}\times7.6^{\prime}$ and binning 2$\times$2 was used during all the observations. Alternated exposures were taken in filters $B$, $V$, $R_c$ and $I_c$ with exposure times of 40, 25, 15 and 15 seconds respectively. A total of 371 target images were acquired during the first night covering an interval of 5.7 hours while 414 images were acquired during the second night for 6.3 hours. Flat field and bias frames were taken during both observing runs.

\begin{figure}
\begin{center}
\includegraphics[scale=0.20]{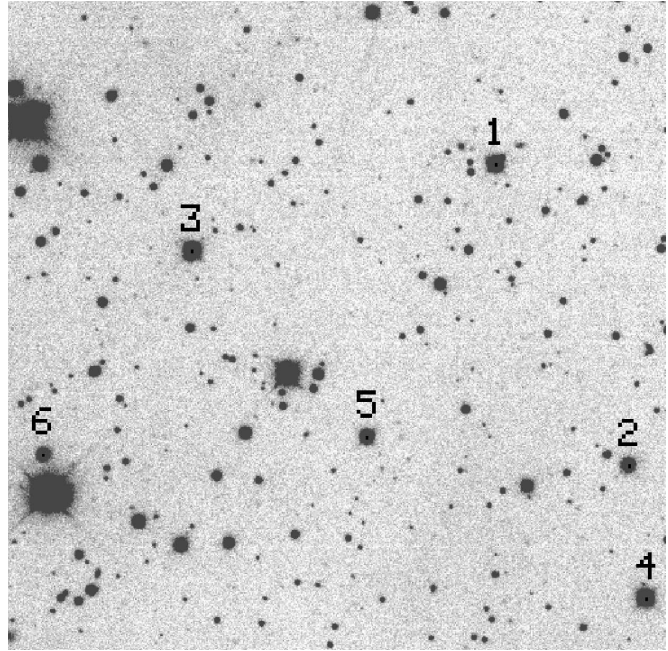}
\end{center}
\caption{Observed field. This finding chart was generated by aligning and adding all the images acquired during the second night of observation. The calibrated $UBV(RI)_c$ magnitudes of the marked stars can be found in Table \ref{tab:UBVRIJHK_mags}.}
\label{fig:field}
\end{figure}

All images were processed using IRAF\footnote{IRAF is distributed by the National Optical Observatories, operated by the Association of Universities for Research in Astronomy, Inc., under cooperative agreement with the National Science Foundation.} routines. Images were bias subtracted and flat field corrected before the instrumental magnitudes of the marked stars in Figure \ref{fig:field} were computed with the standard aperture photometry method. This field was also calibrated in the $UBV(RI)_c$ system and the results, along with the 2MASS magnitudes, are presented in Table \ref{tab:UBVRIJHK_mags}. Based on this information, we decided to use object \#3 as comparison star since it has similar magnitude and color to both V1009 Per and J031642, making differential extinction corrections negligible. Objects \#4, \#5 and \#6 were used as check stars to confirming that the comparison star is not variable. Any part of the data can be provided by the first author upon request.

\begin{sidewaystable}
\caption{$UBV(RI)_c$ and 2MASS magnitudes of the field stars. IDs as in Figure \ref{fig:field}.}
\label{tab:UBVRIJHK_mags}
\begin{center}
{\scriptsize
\begin{tabular}{llcccccccccc}
\hline
ID &  Name & RA (2000) & DEC (2000) & $U$ & $B$ & $V$ & $R_c$ & $I_c$ & $J$ & $H$ & $K_s$ \\
\hline
 1 & V1009 Per              &  49.206375 & 33.504269 & 15.938 & 15.406 & 14.351 & 13.716 & 13.112 & 12.307 & 11.734 & 11.619\\
 2 & CRTS J031642.2+332639  &  49.175527 & 33.444546 & 16.397 & 16.002 & 15.046 & 14.437 & 13.920 & 13.207 & 12.737 & 12.599\\
 3 & LAM260516231           &  49.278550 & 33.487820 & 15.482 & 15.071 & 14.109 & 13.567 & 13.042 & 12.228 & 11.779 & 11.681\\
 4 & LAM260516226           &  49.171755 & 33.418138 & 15.461 & 15.157 & 14.227 & 13.698 & 13.205 & 12.452 & 12.061 & 11.935\\
 5 & 2MASSJ03165701+3327025 &  49.237512 & 33.450702 & 16.434 & 15.955 & 14.987 & 14.458 & 13.988 & 13.186 & 12.837 & 12.680\\
 6 & 2MASSJ03171542+3326522 &  49.314234 & 33.447899 & 16.961 & 16.525 & 15.523 & 14.927 & 14.401 & 13.602 & 13.131 & 13.012\\
\hline
\end{tabular}
}
\end{center}
\end{sidewaystable}

\section{Light elements and orbital period variations of V1009 Per}

The measured times of minimum (ToM), determined by the polynomial fits, of J031642 are presented in Table \ref{tab:minimaJ031642}. These new data permit us to refine the orbital period as:

\begin{equation}
  \label{light_elements2}
HJD(Min.I)=2458136.8210(5)+0^d.2996181(83)\times E,
\end{equation}

\begin{table}
\begin{center}
\caption{CCD (BVRI) times of minima of J031642.}
\label{tab:minimaJ031642}
\scalebox{0.75}{%
\begin{tabular}{ccc}
\hline
HJD	& Epoch & O-C \\
\hline
2458102.665 & -114 & 0	\\
2458136.671 & -0.5 & -0.0005 \\
2458136.822 & 0    & 0.0005 \\
\hline
\end{tabular}}
\end{center}
\end{table}

\begin{figure}
\begin{center}
\includegraphics[scale=0.60]{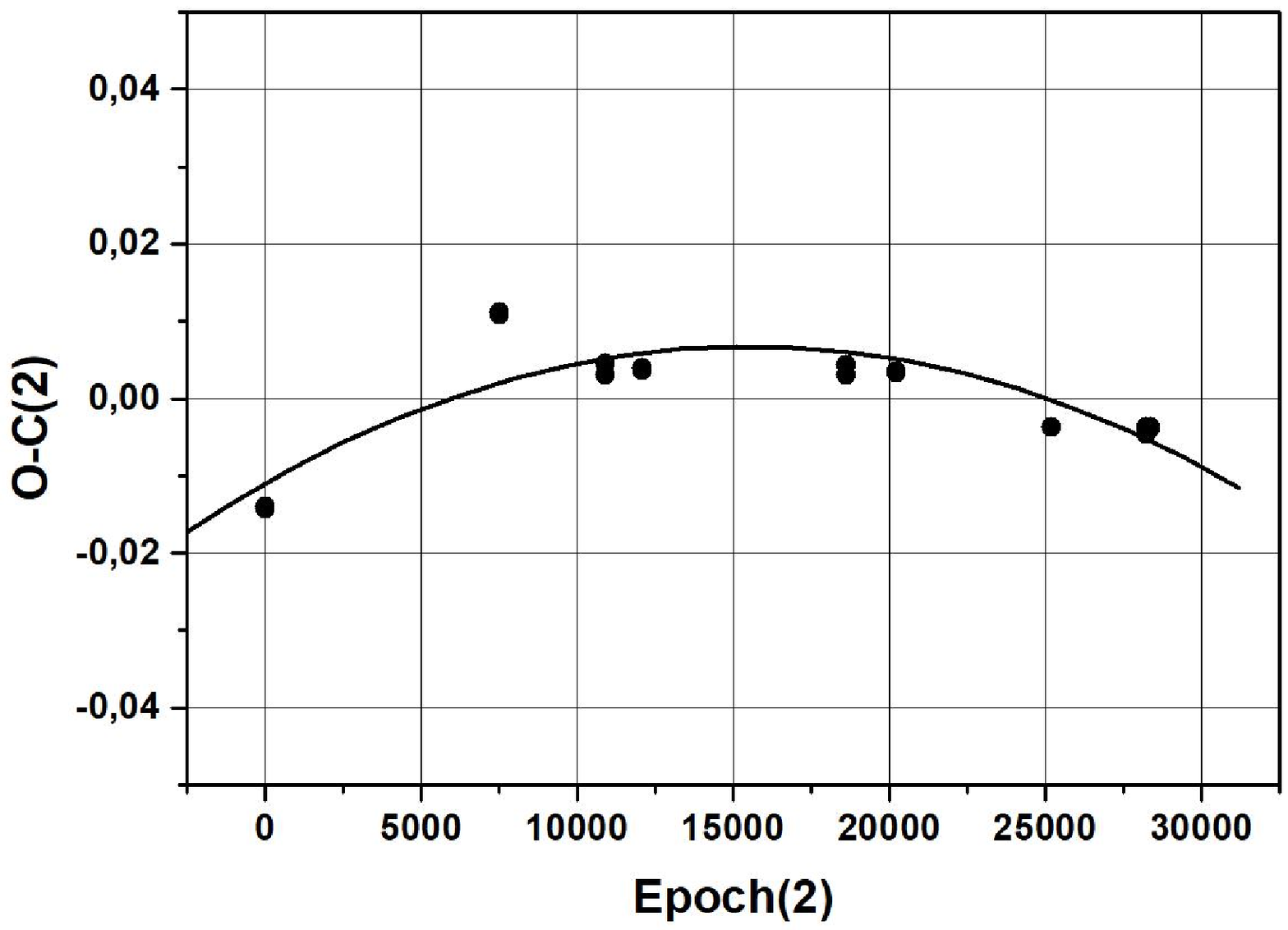}
\end{center}
\caption{O-C diagram of V1009 Per. The solid curve shows the second-order polynomial fit to their data points as given by Eq. \ref{equ:light_elements1}.}
\label{fig:Fig1}
\end{figure}

Based on a careful search for all available eclipsing times of V1009 Per, we collected a total of 16 ToM that, with the new four ToM observed by us (Table \ref{tab:minimaV1009Per}), permit the revision of the ephemerid and the construction of its O-C diagram depicted in Figure \ref{fig:Fig1}.

\begin{table}
\begin{center}
\caption{CCD (BVRI) times of minima of V1009 Per.}
\label{tab:minimaV1009Per}
\scalebox{0.75}{%
\begin{tabular}{lcccl}
\hline
Band   & HJD         & Epoch   & O-C     & Source\\
\hline
Rotse  & 2451491.536 & 0       & -0.0143 & \citet{Kuruslov2011}\\
Rotse  & 2451491.653 & 0.5     & -0.014  & NSVS\\
SWASP  & 2453250.281 & 7511.5  &  0.0111 & SWASP\\
SWASP  & 2453250.397 & 7512    &  0.0109 & SWASP\\
SWASP  & 2454045.87  & 10909.5 &  0.0031 & SWASP\\
SWASP  & 2454045.988 & 10910   &  0.0045 & SWASP\\
SWASP  & 2454320.981 & 12084.5 &  0.0039 & SWASP\\
SWASP  & 2454321.098 & 12085   &  0.0037 & SWASP\\
CCD    & 2455846.852 & 18601.5 &  0.0043 & \citet{Diethelm2012}\\
CCD    & 2455846.968 & 18602   &  0.0031 & \citet{Diethelm2012}\\
CCD(V) & 2456227.909 & 20229   &  0.0034 & \citet{Diethelm2013}\\
CCD    & 2457387.23  & 25180.5 & -0.0037 & Nosal P.\\
BVRI   & 2458102.635 & 28236   & -0.0038 & This paper\\
BVRI   & 2458102.752 & 28236.5 & -0.0046 & This paper\\
BVRI   & 2458136.702 & 28381.5 & -0.0039 & This paper\\
BVRI   & 2458136.819 & 28382   & -0.0037 & This paper\\
\hline
\end{tabular}}
\end{center}
\end{table}

\begin{equation}
  \label{equ:light_elements1}
HJD(Min.I)=2451491.5503(37)+0^d.2341369(2)\times E.
\end{equation}

\begin{figure}
\begin{center}
\includegraphics[scale=0.60]{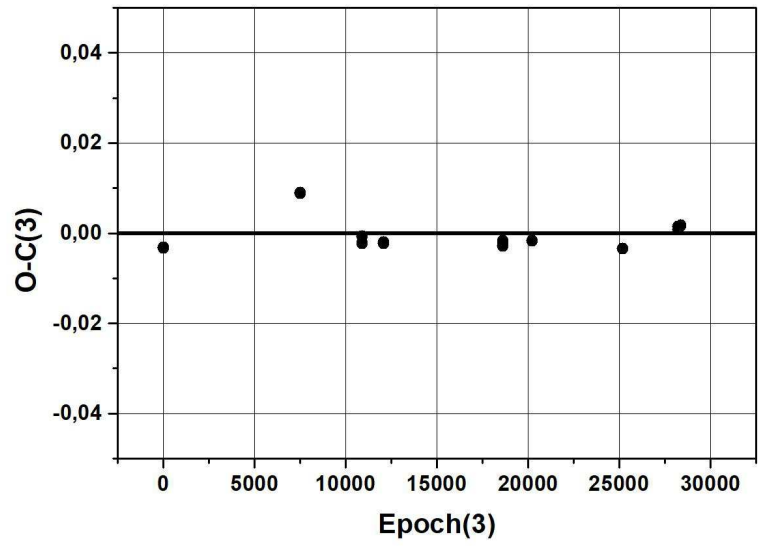}
\end{center}
\caption{O-C diagram of V1009 Per. The solid curve shows the second-order polynomial fit to the data points as given by Eq. \ref{equ:light_elements2}.}
\label{fig:Fig2}
\end{figure}

Applying the ephemeris of Eq. \ref{equ:light_elements1} to those minima, we notice that the trend of the residual has a parabolic shape as shown in Figure \ref{fig:Fig1}. A second order polynomial ephemeris, fitting to all the minima, gives:

\begin{equation}
  \label{equ:light_elements2}
HJD(Min.I)=2451491.5393(28) + 0^d.2341392(4)\times E - 7.4(1.2)\times10^{-11}\times E^2,
\end{equation}

whose residuals behaviour is showed in Figure \ref{fig:Fig2}.

\begin{figure}
\begin{center}
\includegraphics[scale=0.80] {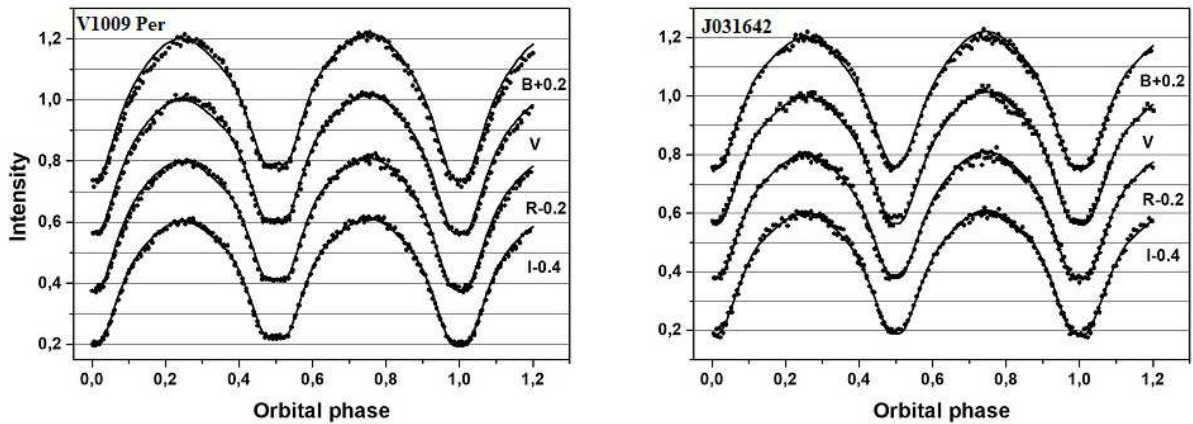}
\end{center}
\caption{CCD B, V R and I light curves of V1009 Per (left) and J031642 (right). Points are the original observations and lines the theoretical light curves with the spot contribution.}
\label{fig:Fig3}
\end{figure}

Since the temporal distribution of the O-C values is rather small (covering only 18 years, spanning nearly 28,400 orbits), the downward parabolic change shown in Figure \ref{fig:Fig1} may be only a part of a long-period cyclic oscillation that may be caused be the presence of a third body. To confirm this conclusion, more times of light minimum are required in the future. 

{From the quadratic term of Eq.(3), it follows that the orbital period may be decreasing at a rate of $\dot{P} = -2.31 \times 10^{-7} days/yr^{-1}$. With the orbital period decrease, the primary component transfer mass to the secondary one, the mass ratio increases and eventually the system evolve into the contact phase.
Period decrease might be caused by mass or angular momentum loss (AML) due to magnetic stellar wind (magnetic breaking) and/or mass transfer from the more massive to the less massive component.}

{According to the formula given by \citet{Bradstreet1994}: $\dot{P}_{AML} = -6.30\times10^{-9} days/yr^{-1}$, this suggests that magnetic breaking is not the main cause of period decrease.
The mass transfer from $M_1$ to $M_2$ or the mass loss from the system can be evaluated using Equations 4 and 5 of \citet{Hilditch2001} for conservative and non-conservative mass loss, respectively. We obtain for V1009 Per $\dot{M}_1 = -1.63\times10^{-7}M_\odot yr^{-1}$ and $\dot{M}_1 = -1.91\times10^{-7}M_\odot yr^{-1}$ for conservative and non conservative mass transfer, respectively.}

{The timescale of the conservative mass transfer ( i.e. the dynamical timescale) can be estimated to be approximately $\tau$dy=  $5.35\times10^{6}$ yrs and $4.59\times10^{6}$ yrs for non conservative mass transfer.}

{On the other hand, the thermal timescale of the massive component can be estimated as $\tau_{th} \approx 3.0\times   10^{7}(M/M_\odot)^{2}(R/R_\odot)^{-1}(L/L_\odot)^{-1}\sim 5.8\times10^{7}$ yrs \citep{Hilditch2001} which is longer than the conservative mass-transfer duration. This suggests that the primary component cannot stay in thermal equilibrium and the mass transfer in V1009 Per is unstable.}

\section{PHOTOMETRIC ANALYSIS USING THE WD CODE}

There are no reported spectroscopic mass ratios for these systems. In order to derive reliable geometric and astrophysical elements, the present observations were analyzed simultaneously using the 2003 (October 2005 revision) version of the Wilson--Devinney (WD) program \citet{Wilson1971}; \citet{Wilson1990}; \citet{Wilson1994}; \citet{Wilson2004}.

We applied the $q$--search method to find the best initial value to be used for $q$ during the light curve analysis. 

From Figure \ref{fig:Fig3} left, it is clearly seen that the light curves of V1009 Per presents a flatter bottom secondary eclipse covering approximately 0.07 in phase; this possibly indicates a total eclipse configuration of the system. 

We used the NASA IPAC database \citep{NASA2015}, interstellar extinction and reddening calculator to compute the $E(B-V)$ value for both the systems which, due to the low Galactic latitude ($-20^\circ$), could be significant.

We have checked the consistence of our determined $(B-V)_0$ values using the period-color relation discovered by Eggen and revised by \citet{Wang1994} as $(B-V)_0=0.062-1.310~logP(days)$. The results are the following:
 
For V1009 Per; NASA's calculator gives a value of $(B-V)_0=0.830$ i.e. $T_1=5280K$, while Wang's equation gives $(B-V)_0=0.736$ i.e. $T_1=5140K$ (difference of $140K$).
For CRTS J031642; NASA's calculator gives $(B-V)_0= 0.736$ i.e. $T_1=5544K$, while Wang's equation gives $(B-V)_0=0.747$ i.e. $T_1=5505K$ (difference of $39K$). The differences between the two values are smaller than the error bars. The $(B-V)$ color values obtained by our observations were corrected with the relative $E(B-V)$ and the resulted values of $(B-V)_0$ were adopted in the determination of the temperature of star 1.

Following \citet{Lucy1967}, the gravity-darkening coefficients of the two components were taken to be 0.32 and bolometric albedo coefficients were set at 0.50 for stars with a convective envelope, \citet{Rucinski1973}. Limb-darkening coefficients of the components were interpolated for square root law from \citet{vanHamme1993} tables.

The shapes of the light curves of these systems are similar to the most frequent light curve shapes of the W UMa-type binary stars, this suggested us to start the W-D analysis directly in Mode 3. Mode 3 in the W-D Code is for over contact binaries (W UMa stars) in which the adjustable parameters used in the Differential Correction calculation are the orbital inclination, $i$, the mean surface effective temperature of the secondary component, $T_2$, the dimensionless surface potentials of the two components, $\Omega_1 = \Omega_2$, and the monochromatic luminosity of the primary component $L_1$.

Due to the common occurrence of third bodies in WUMa systems \citep{Pribulla2006}, the third light was included as an adjustable parameter. The results showed that the values of the third light were negligible and less than the uncertainties. To search for a reliable mass ratio $q$, we made test solutions at the outset using the four light curves in $BVR_cI_c$ colors simultaneously.

\begin{figure}
\begin{center}
\includegraphics[scale=0.62] {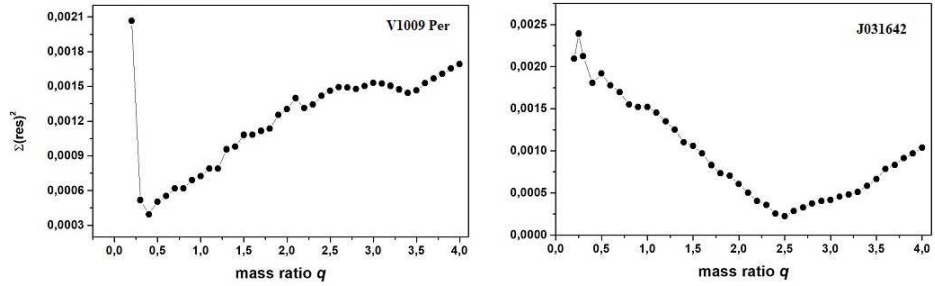}
\end{center}
\caption{The relation $\Sigma(res)^2$ versus mass ratio $q$ in Mode 3 in the WD code for V1009 Per (left) and J031642 (right).}
\label{fig:Fig4}
\end{figure}

The test solutions were computed at a series of assumed mass ratios $q$, with the values from 0.2 to 4 in steps of 0.1 for both the systems, and the behaviour of the sum of squares of residuals, $\Sigma(res)^2$, was used to estimate its value. The relation between the resulting sum of weighted square deviations and $q$ is plotted in Figure \ref{fig:Fig4}. A minimum value was obtained at $q=0.40$ for V1009 Per and $q=2.5$ for J031642. Therefore, we chose the above initial values of mass ratios $q$ and made its an adjustable parameter. Then, we performed a differential correction until it converged and final solutions were derived (Table \ref{tab:LCSols}). It should be noted that the errors of the parameters given in this paper are the formal errors from the WD code and are known to be unrealistically small \citep{Maceroni1997}. For a discussion see \citet{Barani2017}.

\begin{table}
\begin{center}
\caption{Light curves solutions for V1009 Per and J032642. Assumed parameters are marked with $^*$}
\label{tab:LCSols}
\begin{tabular}{lcc}
\hline
           & V1009 Per      & J032642\\
\hline
$i$        & 85.971$\pm$0.145 & 83.245$\pm$0.148\\
$T_1(K)$   & 5280$^*$           & 5544$^*$\\
$T_2(K)$    & 5253$\pm$6       & 5416$\pm$6\\
$\Omega_1=\Omega_2$ & 2.534$\pm$0.004  & 5.872+0.009\\
$q=m_2/m_1$        & 0.362$\pm$0.002  & 0.398$\pm$0.006\\
$A_1=A_2$   & 0.5$^*$            & 0.5$^*$\\
$g_1=g_2$   & 0.32$^*$           & 0.32$^*$\\
$L_{1B}$      & 0.681$\pm$0.002  & 0.318$\pm$0.002\\
$L_{1V}$      & 0.687$\pm$0.001  & 0.311$\pm$0.001\\
$L_{1R}$      & 0.687$\pm$0.001  & 0.308$\pm$0.001\\
$L_{1I}$      & 0.690$\pm$0.001  & 0.307$\pm$0.001\\
$L_{2B}$      & 0.268$\pm$0.002  & 0.634$\pm$0.003\\
$L_{2V}$      & 0.267$\pm$0.002  & 0.639$\pm$0.003\\
$L_{2R}$      & 0.269$\pm$0.002  & 0.645$\pm$0.002\\
$L_{2I}$      & 0.271$\pm$0.002  & 0.654$\pm$0.002\\
$f$        & 0.118$\pm$0.06   & 0.136$\pm$0.04\\
$X_{1B}=X_{2B}$ & 0.749$^*$          & 0.624$^*$\\
$X_{1V}=X_{2V}$ & 0.422$^*$          & 0.319$^*$\\
$X_{1R}=X_{2R}$ & 0.244$^*$          & 0.168$^*$\\
$X_{1I}=X_{2I}$ & 0.131$^*$          & 0.075$^*$\\
$L_3$       & 0              & 0\\
$r_1 (pole)$  & 0.453$\pm$0.001  & 0.289$\pm$0.001\\
$r_1 (side)$  & 0.488$\pm$0.001  & 0.302$\pm$0.001\\
$r_1 (back)$  & 0.591$\pm$0.001  & 0.339$\pm$0.002\\
$r_2 (pole)$  & 0.289$\pm$0.001  & 0.429$\pm$0.001\\
$r_2 (side)$  & 0.303$\pm$0.001  & 0.470$\pm$0.001\\
$r_2 (back)$  & 0.247$\pm$0.003  & 0.499$\pm$0.001\\
\hline
lat spot $(^\circ)$  & 89$\pm$1.8       & 51$\pm$2\\
long spot$(^\circ)$  & 310.1$\pm$3.1    & 270.4$\pm$2.7\\
radius$(^\circ)$     & 24.8$\pm$0.92    & 25.4$\pm$0.88\\
Temp fac.Spot & 1.055$\pm$0.017 & 1.03$\pm$0.02\\
Star       & 2              & 2\\
\hline
Sum (res)2 & 0.00030        & 0.00022\\
\end{tabular}
\end{center}
\end{table}

Results of our analysis confirm that  both the system are shallow contact binaries in good thermal contact. For systems exhibiting high inclination, the mass ratios can be inferred from purely geometric arguments even in the absence of complementary spectroscopic data \citep{Terrell2005}. As shown in Figure \ref{fig:Fig3}, the light curves of both the systems display an inverse O'Connell effect \citep{OConnell1951}. The maximum at phase 0.25 (Max I) is slightly fainter than the one at phase 0.75 (Max II), see Table \ref{tab:DiffHeight}.

\begin{table}
\begin{center}
\caption{Differences in the height of the maxima.}
\label{tab:DiffHeight}
\scalebox{0.75}{%
\begin{tabular}{lcc}
\hline
           & V1009 Per      & J032642\\
\hline
Max II-Max I B & 0.008 & 0.011\\
Max II-Max I V & 0.007 & 0.006\\
Max II-Max I R & 0.016 & 0.008\\
Max II-Max I I & 0.006 & 0.006\\
\end{tabular}}
\end{center}
\end{table}

\begin{figure}
\begin{center}
\includegraphics[scale=0.62] {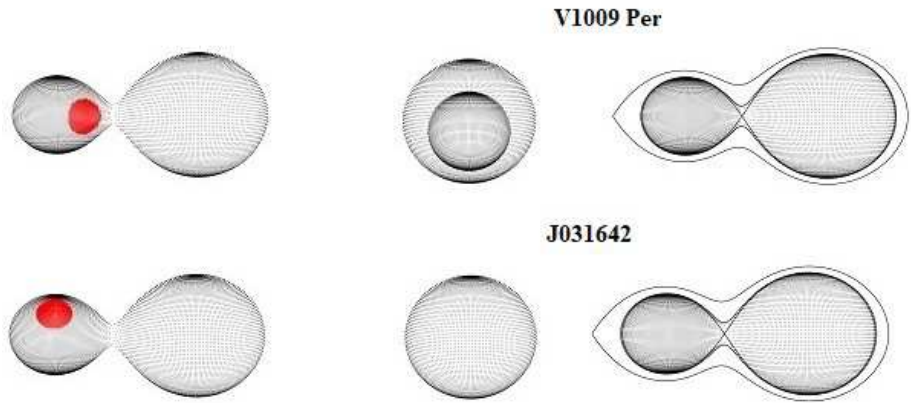}
\end{center}
\caption{Graphic representations of V1009 Per and J031642 according to our solution at the quadrature (left) and at the primary minimum (center). Right, the configuration of the components of the systems in the orbital plane is shown.}
\label{fig:Fig5}
\end{figure}

These features usually indicate wavelength-dependent hot spot activity (rather than a cool spot) on the surface of one component of the system due to the probable impact from mass transfer between the components. The final synthetic light curves calculated by using the whole set of parameters of Table \ref{tab:LCSols} are shown in Figure \ref{fig:Fig3} as continuous lines. The observed and the theoretical light curves are in good agreement. A graphic representations and the Roche geometries of the systems are shown in Figure \ref{fig:Fig5}.

\section{EVOLUTIONARY STATE OF THE SYSTEMS and conclusions}

Since there is no spectroscopic measurement of the orbital elements available presently, the absolute parameters of the system cannot be determined directly. We use the 3D empirical laws of \citet{Gazeas2009b} where the physical parameters of contact binaries are closely correlated with the orbital period and mass ratio. Then using the mass ratio determined from the photometry, we can derive the individual masses and radii.

Luminosities were calculated using the Stefan-Boltzmann law. The physical parameters listed in Table \ref{tab:Elements}, are used to investigate the current evolutionary status of both the systems.

{Recently, \citet{Yildiz2013}, developed a method for computation of initial masses of contact binaries based on stellar modelling with mass loss. Their main assumption is that the mass transfer starts near or after the TAMS phase of the massive component (the progenitor of the secondary component).
Moreover they found that binary systems with initial mass of the secondary, $M_{2i}$, higher than $1.8M_\odot$ become A-subtype and if it is lower than 1.8M$_\odot$; then the systems exhibit W-subtype properties. Applying this method we find $M_{2i} =1.3M_\odot$ for V1009 Per and $M_{2i}=1.4M_\odot$ for J031642. These results agree with the determination of the W-subtype based on the above criterion for the initial mass of the secondary.}

{In Fig. 7, we plotted the components of V1009 Per and J031642 together with other W- and A-type W UMa systems collected by \citet{Yildiz2013} in the logarithmic mass-luminosity (M-L) relations along with the ZAMS and TAMS computed by \citet{Girardi2000}. It is clear that  both components of our systems follow the general pattern of the W-subtype systems and seem to be in good agreement with the well known W-type W UMa systems on the logM-logL plane.}

\begin{table}
\begin{center}
\caption{Estimated absolute elements for V1009 Per and J031642.}
\label{tab:Elements}
\scalebox{0.75}{%
\begin{tabular}{lcc}
\hline
           & V1009 Per      & J032642\\
\hline
  &  Primary star, Secondary star & Primary star, Secondary star\\
Mass ($\mathrm{M}_\odot$) & 0.874$\pm$0.001, 0.317$\pm$0.007 & 1.037$\pm$0.010, 0.414$\pm$0.030\\
Radius ($\mathrm{R}_\odot$) & 0.865$\pm$0.003, 0.474$\pm$0.003 & 1.002$\pm$0.009, 0.661$\pm$0.014\\
Luminosity ($\mathrm{L}_\odot$) & 0.521$\pm$0.004, 0.153$\pm$0.002 & 0.770$\pm$0.010, 0.370$\pm$0.020\\
a ($\mathrm{R}_\odot$) & 1.694$\pm$0.006 & 2.133$\pm$0.029\\
\end{tabular}}
\end{center}
\end{table}

\begin{figure}
\begin{center}
\includegraphics[scale=0.62] {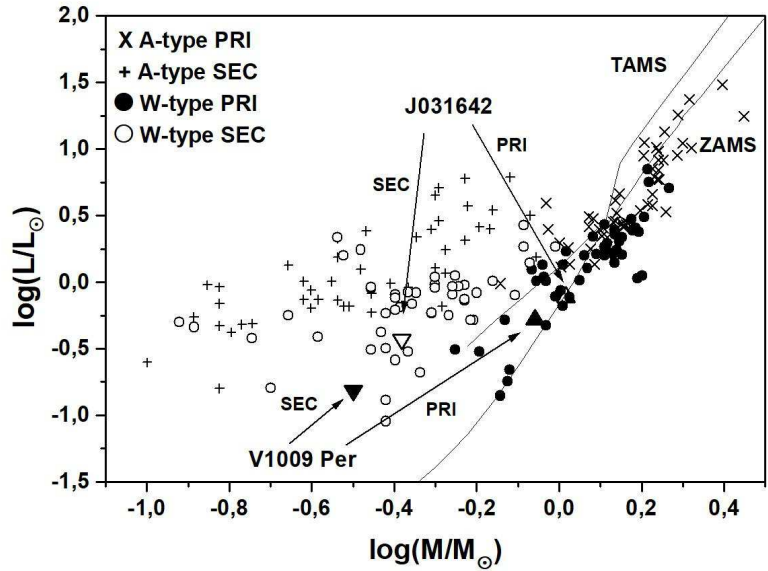}
\end{center}
\caption{Location of the components of V1009 Per and J031642 on the logM - logL diagram. The sample of W UMa type systems was obtained from a compilation of \citet{Yildiz2013} Zero Age Main Sequence (ZAMS) and Terminal Age Main Sequence (TAMS) are taken from \citet{Girardi2000} for the solar chemical composition.}
\label{fig:Fig6}
\end{figure}

In both the systems the light curves exhibit the inverse O'Connell effect with the maximum at phase 0.25 (Max I) slightly fainter than that at phase 0.75 (Max II).

For this reason a hot spot, indicating a probable impact from mass transfer between the components, was placed on the surface of the secondary component.

\acknowledgments

\section*{Acknowledgments}
RM acknowledge the financial support from the UNAM under DGAPA grant PAPIIT IN 100918. This paper is based on observations acquired at the OAN-SPM, Baja California, Mexico.
{This research has made use of the International Variable Star Index (VSX) database, operated at AAVSO, Cambridge, Massachusetts, USA, of the VizieR catalogue access tool, CDS, Strasbourg, France. The original description of the VizieR service was published in A\&AS 143, 23.}

{We acknowledge our anonymous referee for comments that helped to improve this work.}

\end{document}